\documentclass[preprint,showpacs,preprintnumbers,amsmath,amssymb]{revtex4}
\usepackage{graphicx}
\usepackage{dcolumn}
\usepackage{bm}

\newcommand{\be}{\begin{equation}}
\newcommand{\en}{\end{equation}}
\newcommand{\bea}{\begin{eqnarray}}
\newcommand{\ena}{\end{eqnarray}}

\begin{document}


\title{ Tachyon warm inflationary universe models }

\author{Ram\'on Herrera}
\email{ramon.herrera.a@mail.ucv.cl} \affiliation{ Instituto de
F\'{\i}sica, Pontificia Universidad Cat\'{o}lica de
Valpara\'{\i}so, Casilla 4059, Valpara\'{\i}so, Chile.}
\author{Sergio del Campo}
 \email{sdelcamp@ucv.cl}
\affiliation{ Instituto de F\'{\i}sica, Pontificia Universidad
Cat\'{o}lica de Valpara\'{\i}so, Casilla 4059, Valpara\'{\i}so,
Chile.}
\author{Cuauhtemoc Campuzano}
 \email{ccampuz@mail.ucv.cl}
\affiliation{ Instituto de F\'{\i}sica, Pontificia Universidad
Cat\'{o}lica de Valpara\'{\i}so, Casilla 4059, Valpara\'{\i}so,
Chile.}

\date{\today}

\begin{abstract}
 Warm inflationary universe models in a tachyon field theory are studied. General
 conditions required for these models to be realizable are derived and
 discussed. We describe scalar perturbations (in the
 longitudinal gauge) and tensor perturbations for these scenarios. We develop our models
 for a  constant dissipation parameter $\Gamma$ in  one case
 and one  dependent on $\phi$ in  the other case.  We have been successful
 in describing such of
 inflationary universe models.  We  use recent astronomical
 observations  for constraining the parameters appearing in our model. Also, our
 results are compared with their analogous found in the cool
 inflationary case.
\end{abstract}

\pacs{98.80.Cq}
\maketitle

\section{Introduction}

It is well known that many long-standing problems of the Big Bang
model (horizon, flatness, monopoles, etc.) may find a natural
solution in the framework  of the inflationary universe model
\cite{guth,infla}. One of the  success of the inflationary
universe model is that it provides a causal interpretation of the
origin of the observed anisotropy of the cosmic microwave
background (CMB) radiation, and also the distribution of large
scale structures \cite{astros,WMAP}. In standard inflationary
universe models, the acceleration of the universe is driven by a
scalar field (inflaton) with a specific scalar potential, and the
quantum fluctuations associated with this field generate the
density perturbations seeding the structure formation at late time
in the evolution of the universe.  The standard inflationary model
is divided into two regimes; the slow roll and reheating epochs.
In the slow roll period the universe inflates  and all
interactions between the inflation scalar field and any  other
field are typically neglected. Subsequently, a reheating period is
invoked to end  the period of inflation. After reheating, the
universe is filled with radiation \cite{inflat}, and thus  the
universe is connected with the radiation Big-Bang phase.

Warm inflation is an alternative mechanisms for having  successful
inflation and avoiding  the reheating period \cite{warm}. In warm
inflation, dissipative effects are important during inflation, so
that radiation production occurs concurrently  with the
inflationary expansion. The dissipating effect arises from a
friction term which describes the processes of the scalar field
dissipating into a thermal bath via its interaction with other
fields. Also, warm inflation shows how thermal fluctuations during
inflation may  play a dominant role in producing the initial
perturbations. In such of models, the density fluctuations arise
from thermal rather than quantum fluctuations \cite{62526}. These
fluctuations have their origin in the hot radiation and influence
the inflaton through a friction term in the equation of motion of
the inflaton  scalar field \cite{1126}. Among the most attractive
features of these models, warm inflation ends  when the universe
heats up to become radiation domination; at this epoch the
universe stops inflating and "smoothly" enters  a radiation
dominated Big-Bang phase\cite{warm}. The matter  components of the
universe are created by the decay of either the remaining
inflationary field or the dominant radiation field
\cite{taylorberera}.

On the other hand, implications of string/M-theory to
Friedmann-Robertson-Walker cosmological models have recently
attracted  great attention, in particular, those related to
brane-antibrane configurations such as space-like
branes\cite{sen1}. The tachyon field associated with unstable
D-branes might be responsible for cosmological inflation in the
early evolution of the universe, due to tachyon condensation near
the top of the effective scalar potential \cite{sen2} which could
 also add some new form of cosmological dark matter at late times
\cite{Sami_taq}. In fact, historically, as was empathized by
Gibbons\cite{gibbons}, if the tachyon condensate starts to roll
down the potential with small initial velocity, then a universe
dominated by this new form of matter will smoothly evolve from a
phase of accelerated expansion (inflation) to an era dominated by
a non-relativistic fluid, which could contribute to the dark
matter specified above.

Usually, in any  models of warm  inflation, the scalar field which
drives inflation is the standard inflaton field. As far as we
know, a model in which warm inflation is driven  by a tachyonic
scalar field has not been yet studied. The main goal of the
present work is to investigate the possible  realization of a
tachyonic warm inflationary universe model. In this way, we study
cosmological perturbations, which are expressed in terms of
different parameters appearing in our model. These parameters are
constrained from the WMAP three-year data \cite{WMAP}.

In  section \ref{secti}, the dynamics of the tachyon warm
inflationary models is obtained. In  section \ref{pert}, the
cosmological perturbations are investigated. In  section
\ref{exemple}, we use an exponential scalar potential in the high
dissipation regime. There, we distinguish two cases: in the first
scenario, we consider  a constant dissipation rate $\Gamma$, and,
in  the second one, the dissipation rate is considered to be a
function of the tachyonic field, $\phi$.  Here, we study possibles
solutions of the approximated field equations. In section
\ref{exemple2} we develop, in the same approximation, i.e. in the
high dissipation regime, models which are characterized by a power
law scale factor. Finally, in section \ref{conclu}, we give some
conclusions.

\section{Warm Inflationary Model \label{secti}}

As was noted by  Gibbons \cite{gibbons}, a rolling tachyon
condensates  in a spatially flat Friedmann-Robertson-Walker (FRW)
cosmological model is described in terms of an effective fluid
with energy-momentum tensor
$T_{\nu}^{\mu}=diag(-\rho_\phi,p_\phi,p_\phi,p_\phi)$, where the
energy density, $\rho_\phi$, and pressure, $p_\phi$, associated
with the tachyon field are defined  by
\begin{eqnarray}
\rho_\phi=\frac{V(\phi)}{\sqrt{1-\dot{\phi}^2}},\label{1}
\end{eqnarray}
and
\begin{eqnarray}
p_\phi=-V(\phi)\,\sqrt{1-\dot{\phi}^2}\;,
\end{eqnarray}
respectively. Here, $\phi$ denotes the tachyon field (with unit
$1/m_p$, where $m_p$ represents the Planck mass ) and $V(\phi)=V$
is the effective potential associated with this  tachyon field.

 The dynamics of the FRW
cosmological model in the tachyonic case, in the warm inflationary
scenario is described by the equations
 \be
 H^2\,=\kappa \left[\rho_\phi+\rho_\gamma \right]=
 \kappa \left[\frac{V}{\sqrt{1-\dot{\phi}^2}}+\rho_\gamma \right]\label{key_02},
 \en
 \be
 \dot{\rho_\phi}+3H(\rho_\phi+p_\phi)=-\Gamma\dot{\phi}^2\Longrightarrow
\frac{\ddot{\phi}}{(1-\dot{\phi}^2)}+\,3H \;
\dot{\phi}+\frac{V_{,\,\phi}}{V}=-\frac{\Gamma}{V}\sqrt{1-\dot{\phi}^2}\;\;\dot{\phi},
\label{key_01}
 \en
and \be \dot{\rho}_\gamma+4H\rho_\gamma=\Gamma\dot{\phi}^2
,\label{3}\en where  $H=\dot{a}/a$ is the Hubble factor, $a$ is a
scale factor, $\rho_\gamma$ is the energy density of the radiation
field and $\Gamma$ is the dissipation coefficient, with unit
$m_p^5$. The dissipative coefficient  is responsible for the decay
of the tachyon scalar field into radiation during the inflationary
regimen. In general, $\Gamma$ can be assumed as a function of
$\phi$, and  thus $\Gamma=f(\phi)>0$ by the Second Law of
Thermodynamics. Dots mean derivatives with respect to time,
$V_{,\,\phi}=\partial V(\phi)/\partial\phi$,
$\kappa=8\pi/(3m_p^2)$ and we use units in which $c=\hbar=1$.

During the inflationary era the energy density associated with the
tachyonic field is the order of the potential, i.e. $\rho_\phi\sim
V$, and dominates over the energy density associated with the
radiation field, i.e. $\rho_\phi>\rho_\gamma$.  Assuming the set
of slow-roll conditions, $\dot{\phi}^2 \ll 1$, and $\ddot{\phi}\ll
(3H+\Gamma/V)\dot{\phi}$ \cite{warm}, the Friedmann equation is
reduced to
\begin{eqnarray}
H^2&=&\kappa\,V, \label{inf2}
\end{eqnarray}
and  Eq. (\ref{key_01}) becomes
\begin{equation}
3H\left[\,1+r \;\right ]
\dot{\phi}=-\frac{V_{,\,\phi}}{V}\;=-(\ln(V))_{,\,\phi},
\label{inf3}
\end{equation}
where $r$ is the rate defined as
\begin{equation}
 r=\frac{\Gamma}{3H V},\label{rG}
\end{equation}
and parameterizes the  dissipation of our model. For the high (or
weak) dissipation  regimen, $r\gg 1$ (or $r< 1$), i.e. the
dissipation coefficient $\Gamma$ is much greater (or less) than
the product $3\,H \,V$.

If we also considered that  during tachyon warm inflation
radiation production is quasi-stable, then $\dot{\rho}_\gamma\ll 4
H\rho_\gamma$ and $ \dot{\rho}_\gamma\ll\Gamma\dot{\phi}^2$.  From
Eq.(\ref{3}) we obtain that the energy density of the radiation
field becomes
 \begin{equation}
\rho_\gamma=\frac{\Gamma\dot{\phi}^2}{4H}= \sigma T_r^4,\label{rh}
\end{equation}
where $\sigma$ is the Stefan-Boltzmann constant and $T_r$ is the
temperature of the thermal bath. Using  Eqs.(\ref{inf3}),
(\ref{rG}) and (\ref{rh}) we get
\begin{equation}
\rho_\gamma=\sigma\,T_r^4=\frac{r}{12\kappa}\frac{1}{(1+r)^2}\left[\frac{V_{,\,\phi}}{V}\right]^2
.\label{rh-1}
\end{equation}
Introducing the dimensionless slow-roll parameter
\begin{equation}
\varepsilon\equiv-\frac{\dot{H}}{H^2}=\frac{1}{6\kappa}\frac{1}{(1+r)}
\left[\frac{V_{,\,\phi}}{V}\right]^2\frac{1}{V}\;,\label{ep}
\end{equation}
it is possible to find a relation between the energy densities
$\rho_\gamma$ and $\rho_\phi$ given by
\begin{equation}
\rho_\gamma=\frac{r}{2(1+r)}\;\varepsilon\;\rho_\phi.\label{ro}
\end{equation}
Here, the energy density of the tachyonic field during inflation
corresponds to the potential energy, i.e. $\rho_\phi\sim V$.

The second slow-roll parameter  $\eta$ becomes
\begin{equation}
\eta\equiv-\frac{\ddot{H}}{H
\dot{H}}\simeq\frac{1}{3\kappa(1+r)V}\left[\frac{V_{,\,\phi\phi}}{V}-\frac{1}{2}
\left(\frac{V_{,\,\phi}}{V}\right)^2\right].\label{eta}
\end{equation}
The condition under which the tachyon warm inflation epoch could
take place can be summarized with the parameter $\varepsilon$
satisfying   the relation $\varepsilon<1.$ This condition is
analogous  to the requirement in which  $\ddot{a}> 0$. The
condition given above is rewritten in terms of the densities by
using Eq.(\ref{ro}) as
\begin{equation}
\rho_\phi>2\frac{(1+r)}{r} \rho_\gamma,\label{cond}
\end{equation}
which describes the epoch where tachyon warm inflation occur.

On the other hand, inflation ends when the universe heats up, at a
time when $\varepsilon\simeq 1$, which implies
\begin{equation}
\rho_\phi\simeq2\frac{(1+r)}{r} \rho_\gamma.
\end{equation}
The number of e-folds at the end of inflation is given by
\begin{equation}
N(\phi)=-3\kappa\int_{\phi}^{\phi_f}\frac{V^2}{V_{,\,\phi}}(1+r)
d\phi'.\label{N}
\end{equation}

In the following, the subscript $i$ and $f$ are used to denote the
beginning  and the end of  inflation.

\section{The Perturbations  \label{pert}}
In this section we will describe scalar perturbations in the
longitudinal gauge, and then, we will continue describing tensor
perturbations.

\renewcommand{\thesubsection}{\thesection-\arabic{subsection}}

\subsection{Scalar Perturbations  \label{pertadiab}}

By using the longitudinal gauge in the perturbed FRW metric, we
write
\begin{equation}
ds^2=(1+2\Phi)dt^2-a(t)^2(1-2\Psi)\delta_{ij}dx^{i}dx^{j},
\end{equation}
where $\Phi=\Phi(t,{\bf x})$ and $\Psi=\Psi(t,{\bf x})$ are
gauge-invariant variables introduced by Bardeen \cite{Barde}. In
momentum space, for the Fourier components $e^{i{\bf kx}}$, with
$k$ being the wave number, the following set of equations (which
are derived from the perturbed Einstein field equations; we omit
the subscript $k$ here) arises:
\begin{equation}
\Phi=\Psi,
\end{equation}
\begin{equation}
\dot\Phi + H\Phi=\frac{4\pi}{m_p^2}\left[-\frac{4\rho_\gamma\; a\;
v}{3k}+\frac{V\dot{\phi}}{\sqrt{(1-\dot{\phi}^2)}}\;\delta\phi
\right],
\end{equation}
$$
\hspace{-4.5cm}
\frac{(\delta\phi\ddot{)}}{1-\dot{\phi}^2}+\left[3H+\frac{\Gamma}{V}\right](\delta\phi\dot{)}+
\left[\frac{k^2}{a^2}
+(\ln(V))_{,\,\phi\phi}+\dot{\phi}\left(\frac{\Gamma}{V}\right)_{,\,\phi}\right]\;\delta\phi\;\;
$$
\begin{equation}
\hspace{5.cm}=\left[\frac{1}{1-\dot{\phi}^2}
+3\right]\dot{\phi}\;\dot{\Phi}+\left[\dot{\phi}\frac{\Gamma}{V}-2(\ln(V))_{,\;\phi}
\right]\;\Phi,
\end{equation}
\begin{equation}
(\delta\rho_\gamma\dot{)}+4H\delta\rho_\gamma+\frac{4}{3} k
a\rho_\gamma
v-4\rho_\gamma\dot{\Phi}-\dot{\phi}^2\Gamma_{,\,\phi}\delta\phi-\Gamma\dot{\phi}[2(\delta\phi\dot{)}-3\dot{\phi}\Phi]=0,
\end{equation}
and
\begin{equation}
\dot{v}+4H
v+\frac{k}{a}\left[\Phi+\frac{\delta\rho_\gamma}{4\rho_\gamma}+\frac{3\Gamma\dot{\phi}}{4\rho_\gamma}\delta\phi
\right]=0,
\end{equation}
where $v$ appears  from the decomposition of the velocity field
$\delta u_j =-\frac{i a k_j}{k}\;v\;e^{i {\bf kx}} $ $(j=1,2,3)$
\cite{Barde}.

Note that in the case of the scalar perturbations the tachyon and
the radiation fields are interacting. Therefore, isocurvature (or
entropy) perturbations are generated, besides of the adiabatic
ones. This occur because warm inflation can be considered as an
inflationary model with two basics fields \cite{Jora1,Jora}. In
this context, dissipative effects themselves can produce a variety
of spectral ranging between red and blue \cite{62526,Jora}, and
thus producing the running blue to red spectral suggested by WMAP
three-year data\cite{WMAP}. We will return to this point soon.

Since what we need are the non-decreasing adiabatic and
isocurvature modes on large scale $k\ll a H$, (which turn out to
be a weak time dependent quantities), without  loss of generality
we may consistently neglect $\dot{\Phi}$ and those terms
containing two-times derivatives and, combining  with the
slow-roll conditions, the equations for $\Phi$, $\delta\phi$,
$\delta\rho_\gamma$, and $v$ are reduced  to
\begin{equation}
\Phi=\frac{4\pi}{m_p^2}\left(\frac{V\dot{\phi}}{H} \right)\;\left[
1+\frac{\Gamma}{4HV}+\frac{\Gamma_{,\,\phi}\dot{\phi}}{48 H^2 V}
\right]\;\delta\phi,\label{PHI}
\end{equation}
\begin{equation}
\left[3H+\frac{\Gamma}{V}\right](\delta\phi\dot{)}+
\left[(\ln(V))_{,\,\phi\phi}+\dot{\phi}\left(\frac{\Gamma}{V}\right)_{,\,\phi}\right]\;\delta\phi\;=
\left[\dot{\phi}\frac{\Gamma}{V}-2(\ln(V))_{,\;\phi}
\right]\;\Phi,
\end{equation}
\begin{equation}
\delta\rho_\gamma\simeq\frac{\dot{\phi}^2}{4H}[\Gamma_{,\;\phi}\delta\phi-3
\Gamma\Phi]\Longrightarrow\;\frac{\delta\rho_\gamma}{\rho_\gamma}\simeq\frac{\Gamma_{,\;\phi}}{\Gamma}
\;\delta\phi-3\Phi,
\end{equation}
and
\begin{equation}
 v\simeq
-\frac{k}{4aH}\left[\Phi+\frac{\delta\rho_\gamma}{4\rho_\gamma}+
\frac{3\Gamma\dot{\phi}}{4\rho_\gamma}\;\delta\phi\right],
\end{equation}
respectively.

The above equations can be solved taking $\phi$ as the independent
variable instead of $t$. With the help of Eq. (\ref{inf3}) we find
\begin{equation}
\left(3H+\frac{\Gamma}{V}\right)\frac{d}{dt}=\left(3H+\frac{\Gamma}{V}\right)\,\dot{\phi}\frac{d}{d\phi}=
-(\ln(V))_{,\,\phi}\frac{d}{d\phi},
\end{equation}
and  introducing an auxiliary function $\varphi$ given by
\begin{equation}
\varphi=\frac{\delta\phi}{(\ln(V))_{,\,\phi}}\exp\left[\int
\frac{1}{(3H+\Gamma/V)}\left(\frac{\Gamma}{V}\right)_{,\,\phi}\;d\phi\right],\label{solvar}
\end{equation}
we  obtain the following equation for $\varphi$
\begin{equation}
\frac{\varphi_{,\,\phi}}{\varphi}=-\frac{9}{8}\frac{(\Gamma/V+2H)}{(\Gamma/V+3H)^2}
\left[\Gamma+4HV-\frac{\Gamma_{,\,\phi}(\ln(V))_{,\,\phi}}{12H(3H+\Gamma/V)}\right]
\,\frac{(\ln(V))_{,\,\phi}}{V}.\label{var}
\end{equation}
Solving Eq.(\ref{var}) and using Eq.(\ref{solvar}) we find that
\begin{equation}
\delta\phi=C\,(\ln(V))_{,\,\phi} \exp[\Im(\phi)],\label{var2}
\end{equation}
where $\Im(\phi)$ is given by
$$
\hspace{-3.0cm} \Im(\phi)=-\int\left[
\frac{1}{(3H+\Gamma/V)}\left(\frac{\Gamma}{V}\right)_{,\,\phi}+
\frac{9}{8}\frac{(\Gamma/V+2H)}{(\Gamma/V+3H)^2} \left( \Gamma
\frac{^{}}{_{}} \right. \right.
$$
\begin{equation}
\hspace{5.0cm}\left. \left.+\,4HV-\frac{\Gamma_{,\,\phi}
(\ln(V))_{,\,\phi}}{12H(3H+\Gamma/V)}\right)\,\frac{(\ln(V))_{,\,\phi}}{V}\right]d\,\phi,
\end{equation}
and $C$ is a constant of integration.

In this way, the expression for the density perturbations becomes
(see \cite{Liddle})
\begin{equation}
\delta_H=\frac{2}{5}\frac{m_p^2\,\exp[-\Im(\phi)]}{(\ln(V))_{,\,\phi}}\,\delta\phi.\label{33}
\end{equation}
We note here that in the absence of the   dissipation coefficient,
$\Gamma=0$, we see that Eq.(\ref{33}) is reduced to $\delta_H\sim
V\delta\phi/(H\dot{\phi})\sim H\delta\phi/\dot{\phi}$, which
coincides with the expression obtained for cool inflation.

In the case of high dissipation, the dissipation parameter
$\Gamma$ is much greater that the product between the rate
expansion $H$ and the scalar potential, i.e. $r=\Gamma/3HV\gg 1$
and Eq.(\ref{33}) becomes
\begin{equation}
\delta_H^2=\frac{4}{255}\frac{m_p^4\,\exp[-2\;\widetilde{\Im}(\phi)]}{H^2\,r^2}\,\delta\phi^2,\label{331}
\end{equation}
where now $\widetilde{\Im}(\phi):=\Im(\phi)\mid_{r\gg1}$ becomes
\begin{equation}
\widetilde{\Im}(\phi)=-\int\left[ \frac{1}{3H
r}\left(\frac{\Gamma}{V}\right)_{,\,\phi}+ \frac{9}{8}
\left[1-\frac{(\ln\Gamma)_{,\,\phi}(\ln(V))_{,\,\phi}}{36
r\,H^2}\right]\,(\ln(V))_{,\,\phi}\right]d\,\phi.
\end{equation}

In  first approximation, and during the slow-roll phase, the
relation between the density  matter fluctuation, $\delta\rho$,
and the metric perturbation, $\Phi$, is given by

\begin{equation}
\delta\rho\simeq
V_{,\,\phi}\,\delta\phi\simeq\,-2\,[1+r]\,V\,\left[
1+\frac{\Gamma}{4HV}+\frac{\Gamma_{,\,\phi}\dot{\phi}}{48 H^2 V}
\right]^{-1}\,\Phi,\label{r}
\end{equation}
where we have used  Eq. (\ref{PHI}). Note that in the absence of
dissipation, i.e. $\Gamma=0$, we recover the usual relation
$\delta\rho/\rho\simeq\,-2\Phi$ valid for cool inflation, in which
$\rho\simeq V$.

The fluctuations of the tachyon field are generated by thermal
interaction with the radiation field, instead of quantum
fluctuations.  Following Ref.\cite{Bere2}, we may write  in the
case $r\gg 1$,
\begin{equation}
(\delta\phi)^2\simeq\,\frac{k_F\,T_r\,}{2\,m_p^4\,\pi^2},\label{del}
\end{equation}
where  the wave number $k_F$ is defined by $k_F=\sqrt{\Gamma
H/V}=H\,\sqrt{3 r}\geq H$, and corresponds to the freeze-out scale
at which dissipation damps out  the thermally excited
fluctuations. The freeze-out wave number $k_F$ is defined at the
point where the inequality $(\ln(V))_{,\,\phi\,\phi}< \Gamma H/V$,
is satisfied \cite{Bere2}.

Combining Eqs. (\ref{33}), (\ref{331}) and (\ref{del}) we find
\begin{equation}
\delta^2_H\approx
\frac{\sqrt{3}}{75\,\pi^2}\frac{\exp[-2\widetilde{\Im}(\phi)]}{r^{1/2}\,\widetilde{\varepsilon}}\,\frac{T_r}{H},\label{d2}
\end{equation}
or equivalently by using Eqs.(\ref{rh}) and (\ref{rh-1}) we
obtained
\begin{equation}
\delta^2_H\approx\;\frac{\sqrt{3}}{75\,\pi^2}\exp[-2\widetilde{\Im}(\phi)]\,\left[\left(\frac{1}{\widetilde{\varepsilon}}\right)^3\,
\frac{1}{2\,r^2\,\sigma\,\kappa^2\,V} \right]^{1/4},\label{dd}
\end{equation}
where the dimensionless slow-roll parameter in the hight
dissipation period becomes
\begin{equation}
\widetilde{\varepsilon}\approx\frac{1}{6\kappa\,r}\left[\frac{V_{,\,\phi}}{V}\right]^2\frac{1}{V}\;.
\end{equation}

The scalar spectral index $n_s$ is given by
\begin{equation}
n_s -1 =\frac{d \ln\,\delta^2_H}{d \ln k}\label{ns},
\end{equation}
where the interval in wavenumber is related to the number of
e-folds by the relation $d \ln k(\phi)=-d N(\phi)$, and using
Eqs.(\ref{dd}) and (\ref{ns}), it becomes
\begin{equation}
n_s  \approx\,
1\,-\,\left[\frac{3\widetilde{\eta}}{2}+\widetilde{\varepsilon}\left(\frac{2V}{V_{,\,\phi}}\left[2\widetilde{\Im}_{,\,\phi}-
\frac{r_{,\,\phi}}{4r}\right]-\frac{5}{2}
\right)\right],\label{ns1}
\end{equation}
where the second slow-roll parameter  $\eta$ (for $r\gg 1$) is
given by
\begin{equation}
\widetilde{\eta}\approx\frac{1}{3\kappa\,r\,V}\left[\frac{V_{,\,\phi\phi}}{V}-\frac{1}{2}
\left(\frac{V_{,\,\phi}}{V}\right)^2\right].
\end{equation}

One of feature of the three-year data set from WMAP is that it
hints at a significant running in the scalar spectral index
$dn_s/d\ln k=\alpha_s$ \cite{WMAP}. Dissipative effects themselves
can produce a rich variety of spectral ranging between red and
blue \cite{62526,Jora}. From Eq.(\ref{ns1}) we obtain that the
running of the scalar spectral index for our model becomes
$$
\hspace{-1.0cm} \alpha_s=\frac{d n_s}{d\ln k}\approx
\frac{2\;V\;\widetilde{\varepsilon}}{V_{,\;\phi}}\left[
\frac{3\widetilde{\eta}_{,\;\phi}}{2}
+\frac{\widetilde{\varepsilon}_{,\;\phi}}{\widetilde{\varepsilon}}\left(n_s-1+\frac{3\widetilde{\eta}}{2}\right)
+2\widetilde{\varepsilon}\left(\left(\frac{V}{V_{,\;\phi}}\right)_{,\;\phi}\left[2\widetilde{\Im}_{,\,\phi}
-\frac{(\ln r)_{,\;\phi}}{4}\right] \right. \right.
$$
\begin{equation}
\hspace{8.0cm}\left.\left.+\;\left(\frac{V}{V_{,\;\phi}}\right)\left[2\widetilde{\Im}_{,\,\phi\;\phi}
-\frac{(\ln r)_{,\;\phi\;\phi}}{4}\right]
\right)\right].\label{dnsdk}
\end{equation}
In models with only scalar fluctuations, the marginalized value
for the derivative of the spectral index is approximated as
$dn_s/d\ln k=\alpha_s \sim -0.05$ for WMAP-3 only \cite{WMAP}.

In the next section we will study the specific cases in which the
dissipation parameter, $\Gamma$, is either $\Gamma=const.$ or
$\Gamma=\Gamma(\phi)$.

\subsection{ Tensor Perturbations\label{GW}}

As it is mentioned in Ref.\cite{Bha}, the generation of tensor
perturbation during inflation would  produce  stimulated emission
in the thermal background of gravitational wave. This process
changes the power spectrum of the tensor modes through  an extra
temperature dependence   factor $\coth(k/2T)$. The spectrum in
this case is given by
\begin{equation}
A^2_g=\frac{16\pi}{m_p^2}\left(\frac{H}{2\pi}\right)^2\,\coth\left[\frac{k}{2T}\right]\simeq\frac{32\,V}{3\,m_p^4}
\,\coth\left[\frac{k}{2T}\right],\label{ag}
\end{equation}
where the spectral index $n_g$ results as
\begin{equation}
n_g=\frac{d}{d\,\ln k}\,\ln\left[
\frac{A^2_g}{\coth[k/2T]}\right]=-2\,\varepsilon.
\end{equation}
Here, we have used that  $A^2_g\propto\,k^{n_g}\,\coth[k/2T]$ and
Eq.(\ref{ep}) \cite{Bha}.

From  expressions (\ref{d2}), and (\ref{ag}), we may write the
tensor-scalar ration as
$$
R(k_0)=\left.\left(\frac{A^2_g}{P_{\cal
R}}\right)\right|_{\,k=k_0}=\left.\frac{240\sqrt{3}}{25 m_p^2}\,
\left[\frac{(1+r)\,\varepsilon\,H^3}{r^{1/2}\,T_r}\,\exp[2\,\Im(\phi)]
\,\coth\left(\frac{k}{2T}\right)\right]\right|_{\,k=k_0},
$$
where we have used that $\delta_H\equiv\,2\,P_{\cal R}^{1/2}/5$.
For $r\gg 1$, this relation reduces to

\begin{equation}
R(k_0)\;\approx\,\left.\frac{240\sqrt{3}}{25m_p^2}\,
\left[\frac{r^{1/2}\,\widetilde{\varepsilon}\,H^3}{\,T_r}\,\exp[2\,\widetilde{\Im}(\phi)]
\,\coth\left(\frac{k}{2T}\right)\right]\right|_{\,k=k_0},\label{Rk}
\end{equation}
where $k_0$  is referred to as the pivot point.

From the combination of WMAP three-year data\cite{WMAP}  with the
SDSS large scale structure surveys \cite{Teg}, there is found an
upper bound $R(k_0$=0.002 Mpc$^{-1}$)$ <0.28 (95\% CL)$, where
$k_0$=0.002 Mpc$^{-1}$ corresponds to $l=\tau_0 k_0\simeq 30$,
with the distance to the decoupling surface $\tau_0$= 14,400 Mpc.
SDSS measures galaxy distributions at red-shifts $a\sim 0.1$ and
probes $k$ in the range 0.016 $h$ Mpc$^{-1}$$<k<$0.011 $h$
Mpc$^{-1}$. The  recent WMAP three-year results give the values
for  the scalar curvature spectrum $P_{\cal
R}(k_0)\equiv\,25\delta_H^2(k_0)/4\simeq 2.3\times\,10^{-9}$ and
the scalar-tensor ration $R(k_0)=0.095$. These values allow  us to
find  constrains on the  parameters of our model.

\section{Exponential potential in the high dissipation approach.\label{exemple}}

The tachyonic effective potential $V(\phi)$ is one that satisfies
$V(\phi)\longrightarrow$ 0 as $\phi\longrightarrow \infty$.  It
has been argued  that the qualitative tachyonic potential of
string theory can be describe via an exponential potential of the
form \cite{sen2}
\begin{equation}
V(\phi)=V_0 e^{-\alpha\phi},\label{pot}
\end{equation}
where $\alpha$ and $V_0$ are free parameters. In the following we
will take $\alpha> 0$ (with unit $m_p$). The parameter $\alpha$ is
related to the tachyon mass \cite{Fai}. An estimation of these
parameters is give for the cool inflationary case in
\cite{Sami_taq}, where $V_0\sim 10^{-10}m_p^4$ and $\alpha\sim
10^{-6} m_p$.

In the following we will restrict ourselves to the high
dissipation regimen in which $r\gg 1$.

\subsection{ $\Gamma=\Gamma_0=const.$ case\label{GammaConst}}
With $\Gamma=\Gamma_0=const.$ and using the exponential potential
given by Eq.(\ref{pot}), we find that the slow roll parameters
becomes
\begin{equation}
\widetilde{\varepsilon}=\widetilde{\eta}=\frac{1}{6\kappa\;r}\frac{\alpha^2}{V_0
e^{-\alpha\phi}}.\label{eta1}
\end{equation}
The Hubble parameter is given by
\begin{equation}
H(\phi)=\sqrt{\kappa V_0} e^{-\alpha\phi/2},\label{solH}
\end{equation}
while the rate $r$ becomes
\begin{equation}
r=\frac{m_p\Gamma_0}{\sqrt{24\pi}}\;\frac{1}{V_0^{3/2}}\; e^{
\,3\alpha\phi/2}\gg 1.
\end{equation}
 The evolution of the $\dot{\phi}$
during this scenario is governed by the expression
\begin{equation}
 \dot{\phi}=-\frac{V_{,\,\phi}}{3 r H  V},\label{sp2}
\end{equation}
and using Eqs.(\ref{solH})-(\ref{sp2}), we find that the evolution
of the tachyonic field as a function of time is given by
\begin{equation}
\phi(t)=\frac{1}{\alpha}\,\ln\left[e^{\alpha\phi_i}+\frac{\alpha^2
V_0}{\Gamma_0}\,t \right]\label{phisol},
\end{equation}
where $\phi(t=t_i=0)=\phi_i$.  Substituting this last equation
into Eq.({\ref{solH}}), we obtained for the scale factor
\begin{equation}
\frac{a(t)}{a_i}=\exp\left[
\frac{2\Gamma_0}{\alpha^2}\sqrt{\kappa/V_0}\left[
\sqrt{e^{\alpha\phi_i}+\frac{\alpha^2
V_0}{\Gamma_0}\;t}-e^{\alpha\phi_i/2} \right]\right].
\end{equation}
It is not hard to see that  $\ddot{a}(t=t_f)=0$ (or equivalently
$\epsilon_f\simeq 1$ ), where  we get that
\begin{equation}
t_f=\frac{\alpha^2}{4\kappa\,\Gamma_0}-\frac{\Gamma_0\,e^{\alpha\phi_i}}{\alpha^2\,V_0}.\label{tf}
\end{equation}
Thus, $t_f$ represents the time at which inflation ends.
Substituting Eq.(\ref{tf}) into Eq.(\ref{phisol}),we find that the
tachyonic field at the end of inflation becomes
\begin{equation}
\phi_f=-\frac{1}{\alpha}\ln \left[\frac{4
\kappa\,\Gamma_0^2}{\alpha^4 V_0}\right].\label{ff}
\end{equation}

The energy density of the radiation field becomes
\begin{equation}
\rho_\gamma=\frac{\sqrt{6\pi}}{6\kappa}
\left[\frac{\alpha^2}{\Gamma_0}\right]\,V_0^{3/2}\,e^{-3\alpha\phi/2},
\end{equation}
and, in terms of  $\rho_\phi$, it is given by
\begin{equation}
\rho_\gamma=\frac{\sqrt{6\pi}}{6\kappa}
\left[\frac{\alpha^2}{\Gamma_0}\right]\,\rho_\phi^{3/2}.
\end{equation}

Using Eq. (\ref{N}), the total number of e-folds  at the end of
warm inflation results as
\begin{equation}
N_{total}=-3\kappa\int_{\phi_i}^{\phi_f}\frac{V^2}{V_{,\,\tilde{\phi}}}\;r\;
d\tilde{\phi}=\frac{2\Gamma_0}{\alpha^2}\left[\frac{\kappa}
{V_0}\right]^{1/2}[\exp(\alpha\phi_f/2-\exp(\alpha\phi_i/2)],
\end{equation}
where the initial tachyonic field satisfies  $\phi_i<\phi_f$,
since  $V_i>V_f$.

Rewriting  the total number of e-folds  in terms of  $V_f$ and
$V_i$, and  using  Eq.(\ref{ff}), we find
\begin{equation}
N_{total}=\left[1-\left(\frac{V_f}{V_i}\right)^{1/2}\right].\label{N2}
\end{equation}

Note that $V_i>V_f$ and thus  we obtain  $N_{total}\lesssim 1$
e-fold. This problem it is not surprising since this lies in the
slow roll approximation which by virtue  of Eq.(\ref{eta1}) is
valid when $\phi\ll \ln r$, i.e., when $\phi$ is not too large for
the case in which $r\gg 1$. This implies, from Eq.(\ref{phisol}),
that this solution is valid only at early times and therefore it
is not asymptotic. Therefore, the problem of small e-folds is due
to the slow roll approximation. In section V-1, we will  again
consider the case $\Gamma=const.$, but there a power law scale
factor is assumed.

\subsection{ $\Gamma$ as a function of $\phi$.\label{Gammavariable}}

In this case we consider $\Gamma$  as a function of $\phi$.  We
take this to be of the form,
$\Gamma=f(\phi)=c^2\,V(\phi)=c^2\,V_0\,e^{-\alpha\phi}$, where
$c^2> 0$ (with unit of $m_p$). The slow roll parameters are equal,
i.e.
\begin{equation}
\widetilde{\varepsilon}=\widetilde{\eta}=\frac{\alpha^2\,e^{\alpha\phi/2}}{2\;c^2\;(\kappa\,V_0)
^{1/2}},
\end{equation}
and the dissipation parameter $r$ is given  by
\begin{equation}
r=\frac{c^2\,e^{\alpha\phi/2}}{3(\kappa\,V_0)^{1/2}}.
\end{equation}
In this case, we obtain that the evolution of the tachyon field is
given by
\begin{equation}
\phi(t)=\phi_i+\frac{\alpha}{c^2}\,t,
\end{equation}
and the scale factor becomes
\begin{equation}
\frac{a(t)}{a_i}=\exp\left[ \frac{2c^2}{\alpha^2}\sqrt{\kappa
V_0}\,e^{-\alpha\phi_i/2}[1-e^{-\alpha^2\,t/2c^2}] \right].
\end{equation}
The value of the potential at the end of inflation is given by
\begin{equation}
V_f=V_0 e^{-\alpha\phi_f}=\frac{\alpha^4}{4\,\kappa\,c^4},
\end{equation}
and the total number of e-folds  is related to $V_f$ and $V_i$
through
\begin{equation}
V_i=(2\,N_{total}-1)^2\,V_f.
\end{equation}
Note that, unlike in the previous case,  the $N_{total}$ parameter
could assume  appropriate  values (60 or so) for solving the
standard cosmological puzzles.
 To do that,we need the following inequality to be satisfied: $V_i>10^4\,V_f$.

At the beginning of inflation the dissipation parameter  becomes
\begin{equation}
r(\phi)=r_i=\frac{2}{3}\frac{c^4}{(2\,N_{total}-1)\,\alpha^2}\,\gg\,1,
\end{equation}
resulting in the requirement that
$c^2\gg\,\alpha\,(N_{total}-1)^{1/2}$.

On the other hand, from  Eq. (\ref{var2}) we obtain that
\begin{equation}
\delta\phi\simeq\,C\frac{\dot{\phi}}{H}\,\left[1+\frac{c^2}{3H}\right]^{\frac{5}{4}+\frac{21\alpha^2}{16c^4}}\,
\exp\left[\frac{c^2(6H^2(4c^4-21\alpha^2)+Hc^2(8c^4-63\alpha^2)-6c^4\alpha^2)}{32H(c^2+3H)^2}
\right].
\end{equation}
From Eq.(\ref{PHI}), the metric perturbation $\Phi$ becomes given
by
$$
\hspace{-4.0cm}\Phi\simeq-C_1\frac{\dot{H}}{H^2}\left[1+\frac{c^2}{3H}\right]^{\frac{5}{4}+\frac{21\alpha^2}{16c^4}}\,
\left[1+\frac{c^2}{4H}-\frac{\alpha^2\,c^2}{48H^2(c^2+3H)}\right]\,
$$
\begin{equation}
\hspace{4.0cm}\times\;
\exp\left[\frac{c^2(6H^2(4c^4-21\alpha^2)+Hc^2(8c^4-63\alpha^2)-6c^4\alpha^2)}{32H(c^2+3H)^2}
\right],
\end{equation}
where  the zeroth-order term is related the adiabatic mode and the
remaining terms describe the inclusion  of dissipation in the
fluctuation of the scalar field (the entropy mode)\cite{Jora1}.

From Eq.(\ref{ns}) we obtained that the scalar spectral index of
the adiabatic perturbation  becomes
\begin{equation}
n_s\approx1-\left[
\frac{3\,\widetilde{\eta}}{2}-\frac{3}{4}\,\widetilde{\varepsilon}(1+\widetilde{\varepsilon})
\right],
\end{equation}
which becomes $r$ dependent for the case of high dissipation.

From Eq.(\ref{d2}), when $r\gg 1$ or equivalently $c^2\gg 3H$, we
obtain that the scalar power becomes
\begin{equation}
P_{\cal R}(k_0)\approx
\;\frac{\kappa^{19/4}}{2\,\pi^2}\,\left[\frac{c\,}{\alpha^2}\;V(\phi_0)^{5/2}\;T_r\right]\,
\exp\left[\frac{3\alpha^2}{8
c^2\sqrt{\kappa\,V(\phi_0)}}\right],\label{pp}
\end{equation}
and from Eq.(\ref{Rk}) the tensor-scalar ration  is given by
\begin{equation}
R(k_0)\approx\;3\;\pi^2\;\left(\frac{\alpha^2}{c\;\kappa^{11/4}}\right)
\;\left[\frac{1}{V(\phi_0)^{3/2}\;T_r}\right]\,\exp\left[\frac{-3\alpha^2}{8
c^2\sqrt{\kappa\,V(\phi_0)}}\right]\;\coth\left[\frac{k_0}{2T}\right],\label{rr}
\end{equation}
where $V(\phi_0)=V_0\exp[-\alpha\phi_0]$ and  $\phi_0$ represents
the value of the tachyon field when the scale
$k_0=0.002$Mpc$^{-1}$ was leaving the horizon.

Using the WMAP three-year data where $P_{\cal R}(k_0)\simeq
2.3\times 10^{-9}$, $R(k_0)=0.095$ and choosing the parameters
$c^2=10^{10} m_p$, $T\simeq\,T_r\simeq 0.24 \times 10^{16}$ GeV
and $k_0=0.002$Mpc$^{-1}$, we obtained from Eqs.(\ref{pp}) and
(\ref{rr}) that $V(\phi_0)\simeq \,2.43\times 10^{-15} m_p^4$ and
$\alpha\simeq1.76\times 10^{-12}\;m_p$. We noted that the
dissipation coefficient when the scale $k_0$ was leaving the
horizon is the order of $\Gamma(k_0)\sim 10^{-5} m_p^5$.  From
Eq.(\ref{dnsdk}) we found  that it is necessary  to increase the
 value of $\alpha$ by several orders  of magnitude in order to have a running
spectral index $\alpha_s$ close to value given by WMAP
observations.

\section{Power law in the high dissipation approach.\label{exemple2}}

An interesting approach is to choose a scale factor of  power law
type, i.e.  $a\sim t^{p}$, in which $p\geq 2$. This allows us to
get the form of the scalar potential $V(\phi)$ for different
values of $\Gamma$  in the high dissipation approach, i.e.
$\tilde{r}=\frac{\Gamma}{3H\rho_\phi}\gg 1$. Note that if
$\dot{\phi}^2\ll 1$, $\rho_\phi\sim V$ we get
$\tilde{r}\longrightarrow r$.

In the following we will consider  some special cases for the
parameter $\Gamma$.

\subsection{ $\Gamma=\Gamma_0=const.$ case\label{GammaConst2}}

By using  Eqs.(\ref{key_01}) and (\ref{3}), we obtained that
\begin{equation}
\dot{\rho}_\phi+\dot{\rho}_\gamma+4\;H\;\rho_\gamma=0,\label{pw1}
\end{equation}
where we have used for the energy density and pressure associated
with  the  tachyon field the relation
$p_\phi=-\rho_\phi\,(1-\dot{\phi}^2)$. From Eqs.(\ref{key_02}) and
(\ref{pw1}) and choosing  a scale factor of  power law type, we
find that
\begin{equation}
\rho_\gamma=\frac{p}{2\kappa\,t^2}\;,\;\;\;\rho_\phi=\frac{p\;(2p-1)}{2\kappa\,t^2}=\rho_\gamma(2p-1).
\label{den}
\end{equation}
From Eq.(\ref{key_01})(or equivalently from Eq.(\ref{3})) the
temporal evolution of the tachyonic field becomes  given by
\begin{equation}
(\phi-\phi_1)^2=\frac{4(2p-1)}{\kappa\Gamma_0}\,\frac{p}{t},\label{p2}
\end{equation}
where $\phi_1$ is a constant of integration. Substituting
Eq.(\ref{den}) into Eq.(\ref{1}) and using Eq.(\ref{p2}), we
obtained that the potential as a  function of the tachyonic field
$\phi$ becomes
\begin{equation}
V(\phi)=\beta_1\,(\phi-\phi_1)^4\;\sqrt{1-\beta_2\,(\phi-\phi_1)^6}\,,
\end{equation}
 where $\beta_1$ and
$\beta_2$ are given by
$$
\beta_1=\frac{\kappa\Gamma_0^2}{32p\,(2p-1)}\;,\;and\;\;\;\beta_2=\frac{\kappa^2\Gamma_0^2}{64\;p^2\,(2p-1)^2},
$$
respectively.

We  note that Eq. (\ref{pw1}) is valid for
\begin{equation}
\tilde{r}=\frac{\Gamma}{3H\rho_\phi}\gg
1\,\,\Longrightarrow\;\;\frac{128\;p\;(2p-1)^2}{3\kappa^2\;\Gamma_0^2}\gg
(\phi-\phi_1)^6.
\end{equation}

The  number of e-folds becomes
\begin{equation}
N=\int_{t}^{t_f}H(t')\;dt'=2\,p\,\ln\,\left[\frac{\phi-\phi_1}{\phi_f-\phi_1}\right],\label{nto}
\end{equation}
where the  time at the end of inflation is calculated when
$\ddot{a}=0$, or equivalently, by  using the relation
\begin{equation}
[\rho_\phi+\rho_\gamma=-3(p_\phi+p_\gamma)]|_{\phi(t=t_f)}\;,\label{cond2}
\end{equation}
which gives  $\dot{\phi}_f^2=1-(2p+1)/(3(2p-1))$.

From Eq.(\ref{p2})  we find that the time at the end of inflation
 becomes given by
\begin{equation}
t_f=\left[\frac{3\;p\;(2p-1)^2}{4\kappa\Gamma_0
(p-1)}\right]^{1/3}\,,\label{tf}
\end{equation}
and from Eqs.(\ref{nto}) and (\ref{tf}), we have that
\begin{equation}
(\phi-\phi_1)=\left[\frac{4^4\;p^2(2p-1)(p-1)}{3\kappa^2\Gamma_0^2
}\right]^{1/6}\,\exp\left[\frac{N}{2p}\right].
\end{equation}

On the other hand, the spectrum of scalar perturbations becomes
\begin{equation}
P_{\cal
R}(k_0)\approx\frac{2}{\pi^2}\left[\frac{V(\phi_0)^{23}}{\beta_1^{37/2}}\right]^{1/4}
\left(\frac{T_r}{(\phi_0-\phi_1)^{13}\;[4-7\beta_2(\phi_0-\phi_1)^6]^2}\right)\,,\label{pp1}
\end{equation}
and  the tensor-scalar ration is given by
\begin{equation}
R(k_0)\approx
\frac{512\pi}{m_p^2}\left[\frac{\beta_1^{53/2}}{V(\phi_0)^{23}}\right]^{1/4}\,
\left(\frac{(\phi_0-\phi_1)^{17}\;[4-7\beta_2(\phi_0-\phi_1)^6]^2}
{\Gamma_0^2\,T_r}\right)\,\;\coth\left[\frac{k_0}{2T}\right].\label{rr1}
\end{equation}

Using the WMAP three-year data, in which $P_{\cal R}(k_0)\simeq
2.3\times 10^{-9}$, $R(k_0)=0.095$ and choosing the parameters
$p=2$, $T\simeq\,T_r\simeq 0.24 \times 10^{16}$ GeV and
$k_0=0.002$Mpc$^{-1}$, we obtain from Eqs.(\ref{pp1}) and
(\ref{rr1}) that $(\phi_0-\phi_1)\approx 2\times 10^{3}m_p$ and
$\Gamma_0\approx 10^{-11}m_p^5$. Here the scalar potential when
the scale is $k_0=0.002$Mpc$^{-1}$ is leaving the horizon becomes
$V(\phi_0)\sim 3\times 10^{-10}m_p^4$. Note that this value of
$V(\phi_0)$ becomes similar to that found for cool inflation where
a chaotic potential is used \cite{Bha}. Another case of interest
is that of $p=4$, where it is found that $(\phi_0-\phi_1)\approx
10^{7}m_p$, $\Gamma_0\approx 10^{-22}m_p^5$ and $V(\phi_0)\sim
 10^{-17}m_p^4$.  In this case, from Eq.(\ref{dnsdk})
 we found that if we decrease the value of
 $(\phi-\phi_1)$ by a few orders  of magnitude, then the
 running spectral index $\alpha_s$ takes
  a value which agrees with the value specified  by the WMAP observations.

\subsection{ $\Gamma$ as a function of t \label{GammaConst3}}

In this case, for simplicity we considered  that the dissipation
coefficient is $\Gamma=\gamma/t$, where $\gamma$ is a
positive-definite constant ( with the unit of $m_p^4$). By using
Eqs.(\ref{key_01}) and (\ref{den}), we obtain  that the tachyonic
field becomes
\begin{equation}
(\phi-\tilde{\phi}_1)=\beta_3\,\ln[t],\label{lnt}
\end{equation}
where $\tilde{\phi}_1$ is a constant of integration and
$\beta_3=\sqrt{(2p-1)\;p/\kappa\gamma}$. We also note that Eq.
(\ref{pw1}) is valid for
\begin{equation}
\tilde{r}=\frac{\Gamma}{3H\rho_\phi}\gg
1\,\,\Longrightarrow\;\;\frac{2\kappa\gamma}{3\,p^2\;(2p-1)}\gg
\exp(-2(\phi-\phi_1)/\beta_3).
\end{equation}

 The potential is now  given
by
\begin{equation}
V(\phi)=\left(\frac{\gamma\;\beta_3^2}{2}\right)\,\exp\left[\frac{-2(\phi-\tilde{\phi}_1)}{\beta_3}\right]\,
\sqrt{1-\beta_3^2\,\exp\left[\frac{-2(\phi-\tilde{\phi}_1)}{\beta_3}\right]}.
\end{equation}
From Eqs.(\ref{cond2}) and (\ref{lnt}), we obtain that the  time
at the end of inflation is given by
\begin{equation}
t_f=\frac{(2p-1)}{2}\;\sqrt{\frac{3\,p}{\kappa\;\gamma\;(p-1)}}\;.
\end{equation}
The tachyonic field in terms of the $N$ e-folds parameter becomes
 \begin{equation}
(\phi-\tilde{\phi}_1)=\beta_3\left[\ln\left(\frac{(2p-1)}{2}\;
\sqrt{\frac{3\;p}{\kappa\;\gamma\;(p-1)}}\right)-\frac{N}{p}\right].
\end{equation}
Following a scheme similar  to those of the previous cases, in
order to calculate the spectrum of scalar perturbations and the
tensor-scalar ratio, we obtained  that
$(\phi_0-\tilde{\phi}_1)\approx 69.6 m_p$ and $\gamma\approx
0.02\; m_p^4$, and  the potential when the scale
$k_0=0.002$Mpc$^{-1}$ is leaving the horizon  becomes
$V(\phi_0)\sim 0.2\times 10^{-14}\, m_p^4$. We have chosen  the
values $p=2$ and $T\simeq\,T_r\simeq 0.24 \times 10^{16}$ GeV. For
the value of the  parameters given above, and  from
Eq.(\ref{dnsdk}),  we find that the running spectral index
$\alpha_s$  lies in  the range established by WMAP observations.

\section{Conclusions \label{conclu}}

In this paper we have investigated the tachyonic warm inflationary
scenario. In the slow roll approximation we have found a general
relationship between the radiation and tachyon energy densities.
This has led us to a general criterion   for tachyon warm
inflation to occur; see Eq.(\ref{cond}).

In relation to the corresponding perturbations, the contributions
of the adiabatic and entropy modes were obtained explicitly. Here,
it is shown that the  dissipation parameter  plays a crucial role
in producing the  entropy mode. Additionally, we have found a
general relation for the density perturbation expressed by
Eq.(\ref{33}). The tensor perturbation are generated via
stimulated emission into the existing thermal background (see
Eq.(\ref{ag})), and the tensor-scalar ratio is modified by a
temperature dependent factor.

Our first specific models are described using  an exponential
scalar potential and for two different  dependences     of the
dissipation coefficient, $\Gamma$. In the first case, we took
$\Gamma=\Gamma_0=Cte.$ and we have found that the number of
e-folds in this case  is insufficient. For the case in which the
dissipation coefficient $\Gamma$  is taken to be a function of the
tachyon field, i.e. $\Gamma=f(\phi)=c^2\;V(\phi)$, it has been
possible to describe appropriately  tachyon warm inflationary
universe models. Also, in  place of choosing a specific form of
the scalar tachyon potential a priory, we have assumed a power law
for the scale factor. Here, we have described inflationary models
for $\Gamma = \Gamma_0=const$ and  for $\Gamma$ variable. In both
cases, we have obtained explicit expressions for the corresponding
scalar potential. By using the WMAP three-year data, we have found
some constraints for the parameters appearing in our model. From
the normalization of the WMAP data, the potential becomes of the
order of $V(\phi_0)\sim 10^{-15}m_p^4$ when it leaves the horizon,
at the scale of $k_0=0.002$Mpc$^{-1}$.

Dissipative effects plays a crucial role in producing the entropy
mode; they can themselves  produce a rich variety of spectral
ranging between red and blue. The possibility of a spectrum which
runs from blue to red is particularly interesting because it is
not commonly seen in inflationary models, which typically predict
red spectral. Models of inflation with dissipative effects and
models with interacting fields have much more freedom to yield
spectral. In particular, we anticipate that the Planck mission
will significantly enhance our understanding of $\alpha_s$ by
providing hight quality measurements of the fundamental power
spectrum over a large wavelength range  than WMAP.
 Summarizing, we
have been successful in describing  tachyon warm inflationary
models for describing the early epoch of the universe in the high
dissipation regimen.

\begin{acknowledgments}
R. H. was supported by the ``Programa Bicentenario de Ciencia y
Tecnolog\'{\i}a" through the Grant ``Inserci\'on de Investigadores
Postdoctorales en la Academia" \mbox {N$^0$ PSD/06}. S. d. C. was
supported by COMISION NACIONAL DE CIENCIAS Y TECNOLOGIA through
FONDECYT grants N$^0$ 1030469, N$^0$1040624 and N$^0$1051086.
Also, from UCV-DGIP N$^0$ 123.764. C. C. was supported by
MINISTERIO DE EDUCACION through Postdoctoral MECESUP Grants FSM
0204.
\end{acknowledgments}


\end{document}